\documentclass[12pt]{iopart}

\usepackage{iopams}
\usepackage{amstext}
\usepackage{graphics}
\usepackage{epsfig}
\usepackage{citesort}

\def\Bid{{\mathchoice {\rm {1\mskip-4.5mu l}} {\rm
{1\mskip-4.5mu l}} {\rm {1\mskip-3.8mu l}} {\rm {1\mskip-4.3mu l}}}}

\newcommand{\bra}[1]{{\langle{#1}|}}
\newcommand{\ket}[1]{{|{#1}\rangle}}
\newcommand{\A}{\hat{a}}
\newcommand{\C}{\hat{a}^\dagger}

\begin{document}


\title{$SU(N)$-symmetric quasi-probability distribution functions}
\author{Todd Tilma}
\address{National Institute of Informatics, 2-1-2 Hitotsubashi, Chiyoda-ku, Tokyo 101-8430, Japan}
\author{Kae Nemoto}
\address{National Institute of Informatics, 2-1-2 Hitotsubashi, Chiyoda-ku, Tokyo 101-8430, Japan}
\eads{\mailto{ttilma@nii.ac.jp}, \mailto{nemoto@nii.ac.jp}}


\begin{abstract}
We present a set of $N$-dimensional functions, based on generalized $SU(N)$-symmetric coherent states, that represent finite-dimensional Wigner functions, Q-functions, and P-functions.
We then show the fundamental properties of these functions and discuss their usefulness for analyzing $N$-dimensional pure and mixed quantum states.
\end{abstract}


\maketitle
\eqnobysec

\section{Introduction}\label{Introduction}
Recent developments in quantum technology have brought us a capability of manipulating and measuring a quantum system larger than two qubits using a number of different physical systems.  
The quantum states of such high-dimensional systems have been experimentally measured and characterized~\cite{PhysRevLett.86.4435}.
The standard method to evaluate quantum states in experiments is state tomography.  
With state tomography, we can reconstruct the density matrix of the system.  
Since the density matrix contains all the information of the quantum state we have, we are able in principal to calculate any characteristic of the system.
The only problem is that the larger system gets, the exponentially more elements we have to measure and compute to characterize the system, and the analysis of the quantum nature of a state quickly becomes intractable.  

Fortunately, there has been some recent progress in state characterization and visualization. 
A tomographic method for large systems with certain symmetries~\cite{Eisert}, and a state visualization method for discrete systems that extends the Wigner function~\cite{Wootters19871,PhysRevA.70.062101} have been developed.
By contrast, for qunats (continuous variables) a long history exists for establishing a tool set that efficiently represents and analyzes quantum states in an infinite-dimensional Hilbert space.  
In particular, the standard method in quantum optics is based on mapping operator functions to corresponding $c$-number functions.  
For example, the Wigner function~\cite{Wigner1932,Moyal1949,Ozizmir1967,Leaf1968-1,Leaf1968-2}, Q-function~\cite{Husimi1940}, and P-function~\cite{PhysRevLett.10.277,PhysRev.131.2766}, which have been widely used in both theoretical and experimental analysis, are all $c$-number functions using coherent states.  
The coherent state is defined as ($\ket{\alpha} = D(\alpha,\alpha^*) \ket{0} = \rme^{\alpha \C - \alpha^* \A} \ket{0}$)~\cite{Klauder-Sudarshan}.  
Here, $D(\alpha,\alpha^*)$ is a displacement operator in the phase space where $\ket{0}$ is the vacuum.  
It can also be taken as the kernel that generates the Wigner function; hence~\cite{Glauber1969}
\begin{equation}
\label{Optical-Displacement-CF}
\mathrm{W}(\alpha,\alpha^{*}) = \Tr[\rho \cdot 2 D(\alpha,\alpha^*) (-1)^{\C \A} D^{-1}(\alpha,\alpha^*)],
\end{equation}
Similarly, the Q-function and P-function can be obtained through~\eref{Optical-Displacement-CF} by replacing $D(\alpha,\alpha^*)$ with the appropriate operator, respectively.

In this paper, we expand the use of these functions to general $SU(N)$ systems by using $SU(N)$-symmetric coherent states. 
To begin, $SU(2)$ coherent states for $d$-level systems can be generalized as the trajectory of the $SU(2)$ group action $U_{2}^{M/2}$ on the lowest wight state $\ket{\psi_{0}}$~\cite{Perelomov,PerelomovB}; hence 
\begin{eqnarray}
\label{coherent-state-atomic}
\fl
\ket{\theta, \phi} &=& U_{2}^{M/2}(\theta,\phi) \ket{\psi_{0}}  \nonumber \\
\fl
&=&\sum_{m=-\frac{M}{2}}^{\frac{M}{2}} \left( \begin{array}{c} M \\ \frac{M}{2}+m \end{array} \right)^{\frac{1}{2}} \, \sin\biggl[\frac{\theta}{2}\biggr]^{\frac{M}{2}+m} \cos\biggl[\frac{\theta}{2}\biggr]^{\frac{M}{2}-m} \rme^{\rmi(\frac{M}{2}-m)\phi} \ket{\frac{M}{2},m}.
\end{eqnarray}
Here an Euler angle decomposition was used for $U_{2}^{M/2}(\theta,\phi)$ and $d=M+1$, which is usually denoted as $d=2j+1$ where $j$ is a quantum number.  
We introduced the integer parameter $M$ only for convenience in the generalization later in this paper.   

With the $SU(2)$ coherent state given in~\eref{coherent-state-atomic}, the Wigner function, Q-function, and P-function have been defined~\cite{PhysRevA.6.2211,PhysRevA.49.4101}.  
They are successfully used to analyze atoms in a trap~\cite{PhysRevA.49.4101} and spin-squeezed states of two-component Bose-Einstein condensates~\cite{0305-4470-34-10-309,nature08988}.  
However, the main obstacle in using this state representation is the difficulty in adopting its composite structure into the analysis of the quantum system. 
For example, in quantum information processing, it is important to maintain the properties dependent on the composite structure of the system; quantities such as entanglement only have meaning with it.  
To accommodate a more detailed system structure, we first need to generalize the state representation method to $SU(N)$ systems.  

Our starting point is to generalize~\eref{coherent-state-atomic} to $SU(N)$.  
This generalization can be done as~\cite{Nemoto2000} 
\begin{equation}
\label{TheCoherentStates}
\ket{(\boldsymbol{\theta},\boldsymbol{\phi})_{N}^{M}} = U_{N}^{M}(\boldsymbol{\theta},\boldsymbol{\phi}) \ket{\psi_{0}}.
\end{equation}
Here, $\boldsymbol{\theta} \equiv \theta_1, \theta_2, \ldots, \theta_{N-1}$, $\boldsymbol{\phi}\equiv \phi_{1}, \phi_{2}, \ldots, \phi_{N-1}$, and $\ket{\psi_{0}}$ is the lowest weight state in terms of the group operation $U_{N}^{M}$.  
The quantum number $M$ defines the dimension $d$ of the representation, hence the system size is given as 
\begin{equation}
\label{TheOmega}
d=b_{N,M} \equiv \left( \begin{array}{c} N+M-1 \\ M \end{array} \right).
\end{equation}
We will revisit this and explicitly define the coherent states with an Euler angle parametrization~\cite{Tilma2} in Section~\ref{GenCS}.

Using $SU(3)$ coherent states, a Winger function has recently been constructed~\cite{Luis-495302}.  
More general $SU(N)$-symmetric distribution functions have been shown to exist~\cite{Braunstein-1004.5425,Klimov-1008.2920}.  
Furthermore, the Q-function can be generalized rather straightforwardly with these coherent state, however no general quasi-probability functions are constructed using the coherent states defined in~\eref{TheCoherentStates}.  
In this paper, we construct Wigner and P-functions for $SU(N)$ in both the $M=1$ and $2$ cases.

This paper is organized as follows. 
We first review the construction and properties of our generalized coherent states and then show how they help build a $SU(N)$-symmetric Wigner function, Q-function, and P-function through the Stratonovich-Weyl correspondence~\cite{Stratonovich56}.
We then discuss some general properties of the functions and then conclude with an example.

\section{$SU(N)$-symmetric coherent states}\label{GenCS}
We start by explicitly defining our $SU(N)$ coherent states for $d$-dimensional systems, where $d$ is given in~\eref{TheOmega}.  
First we denote the $SU(N)$ generators~\cite{Nemoto2000} by the set 
\begin{equation}
\label{TheLambdas}
\{ \Lambda_{N,M}(k) \} \; \text{where } k=1,2,\ldots,N^{2}-1.
\end{equation}
This set is made up of off-diagonal generators
\begin{equation}
\Lambda_{N,M}^{\{1\}}(a,b) \text{ and } \Lambda_{N,M}^{\{2\}}(a,b) \text{, for } a,b = 1,2,3,\ldots,N;\, a < b,
\end{equation} 
and diagonal generators 
\begin{equation}
\Lambda_{N,M}^{\{3\}}(c) \text{ for } 1 \leq c \leq N-1.
\end{equation} 
A detailed procedure to construct these matrices and their properties is given in~\ref{Lambda}. 
When $d=N$, i.\ e.\ $M=1$, the representation is fundamental and the generators above reduce to the generalized Gell-Mann matrices $\{ \lambda_{k} \}$ for $SU(N)$~\cite{Greiner,Georgi}.  
In particular, when $d=3$, i.\ e.\ $M=1$ and $N=3$, this procedure generates the standard Gell-Mann matrices $\{ \lambda_{k} \}$ for $SU(3)$:
\begin{equation}
\label{SU3-lambdas}
\begin{array}{crcr}
\fl
\Lambda_{3,1}^{\{1\}}(1,2) \equiv \Lambda_{3,1}(1) = \left( \begin{array}{cccc}
                     0 & 1 & 0 \\
                     1 & 0 & 0 \\
                     0 & 0 & 0 \end{array} \right), \,
\Lambda_{3,1}^{\{2\}}(1,2) \equiv \Lambda_{3,1}(2) = \left( \begin{array}{crcr} 
                     0 & -\rmi & 0 \\
                     \rmi &  0 & 0 \\
                     0 &  0 & 0 \end{array} \right), \\
\fl
\Lambda_{3,1}^{\{1\}}(1,3) \equiv \Lambda_{3,1}(4) = \left( \begin{array}{crcr} 
                     0 & 0 & 1 \\
                     0 & 0 & 0 \\
                     1 & 0 & 0 \end{array} \right), \,
\Lambda_{3,1}^{\{2\}}(1,3) \equiv \Lambda_{3,1}(5) = \left( \begin{array}{crcr} 
                     0 & 0 & -\rmi \\
                     0 & 0 &  0 \\
                     \rmi & 0 &  0 \end{array} \right), \\
\fl
\Lambda_{3,1}^{\{1\}}(2,3) \equiv \Lambda_{3,1}(6) = \left( \begin{array}{crcr} 
                     0 & 0 & 0 \\
                     0 & 0 & 1 \\
                     0 & 1 & 0 \end{array} \right), \,
\Lambda_{3,1}^{\{2\}}(2,3) \equiv \Lambda_{3,1}(7) = \left( \begin{array}{crcr} 
                     0 & 0 &  0 \\
                     0 & 0 & -\rmi \\
                     0 & \rmi &  0  \end{array} \right), 
\end{array}    
\end{equation}
and
\begin{equation}
\label{SU3-CSA}
\begin{array}{crcr}
\fl
\Lambda_{3,1}^{\{3 \}}(3) \equiv \Lambda_{3,1}(3) = \left( \begin{array}{crcr} 
                     1 &  0 & 0 \\
                     0 & -1 & 0 \\
                     0 &  0 & 0 \end{array} \right), \,
\Lambda_{3,1}^{\{3 \}}(8) \equiv \Lambda_{3,1}(8) = \frac{1}{\sqrt{3}}\left( \begin{array}{crcr} 
                     1 & 0 &  0 \\
                     0 & 1 &  0 \\
                     0 & 0 & -2 \end{array} \right).
\end{array}
\end{equation}

Given the generators $\{ \Lambda_{N,M}(k) \}$, we can employ the parametrization given in~\cite{Tilma2} for $SU(N)$.  
Using this parametrization, a $SU(N)$ operator for $M$ can be decomposed as
\begin{eqnarray}
\label{eq:suN} 
U_{N}^{M}(\boldsymbol{\theta},\boldsymbol{\phi}) = \biggl(\prod_{N \geq z \geq 2} \, \prod_{2 \leq y \leq z} A_{M}(y,j(z))\biggr) \times B_{M},
\end{eqnarray}
where
\begin{eqnarray}
A_{M}(y,j(z)) = \rme^{\rmi \Lambda_{N,M}^{\{3 \}}(3) \phi_{(y-1)+j(z)}} \rme^{\rmi \Lambda_{N,M}^{\{2\}}(1,y) \theta_{(y-1)+j(z)}},
\end{eqnarray}
and
\begin{eqnarray}
B_{M} = \prod_{1 \leq c \leq N-1} \rme^{\rmi \Lambda_{N,M}^{\{3 \}}((c+1)^2-1) \phi_{(N(N-1)/2) + c}}.
\end{eqnarray}
Here $j(z) =0$ for $z = N$ and $j(z)=\sum_{i=1}^{N-z}(N-i)$ for $z \neq N$.  
The $A_{M}(y,j(z))$ terms are from the off-diagonal generators and the $B_{M}$ term is from the diagonal generators.
For example, for $SU(3)$,~\eref{eq:suN} gives us~\cite{MByrd1}
\begin{eqnarray}
\label{su3eas}
\fl
U &=& U_{3}^{M}(\boldsymbol{\theta},\boldsymbol{\phi}) \\
\fl 
&=& \rme^{\rmi \Lambda_{3,M}^{\{3 \}}(3) \phi_1} \rme^{\rmi \Lambda_{3,M}^{\{2\}}(1,2) \theta_1} \rme^{\rmi \Lambda_{3,M}^{\{3 \}}(3) \phi_2} \rme^{\rmi \Lambda_{3,M}^{\{2\}}(1,3) \theta_2} \rme^{\rmi \Lambda_{3,M}^{\{3 \}}(3) \phi_3} \rme^{\rmi \Lambda_{3,M}^{\{2\}}(1,2) \theta_3} \rme^{\rmi \Lambda_{3,M}^{\{3 \}}(3) \phi_4} \rme^{\rmi \Lambda_{3,M}^{\{3 \}}(8) \phi_5}. \nonumber
\end{eqnarray}

Using~\eref{eq:suN}, the generalized coherent state for $SU(N)$ in the fundamental representation $\ket{(\boldsymbol{\theta},\boldsymbol{\phi})_{N}^{1}}$ can explicitly be written as
\begin{equation}
\label{TheCS-Spin1}
\fl
\ket{(\boldsymbol{\theta},\boldsymbol{\phi})_{N}^{1}} = \varphi 
\left[\begin{array}{c}
\rme^{\rmi (\phi _{1} + \phi _{2} + \cdots + \phi _{N-2} + \phi _{N-1})}\cos[\theta_1]\cos[\theta_2] \cdots \cos[\theta_{N-2}] \sin[\theta_{N-1}] \\
-\rme^{\rmi (-\phi _{1} + \phi _{2} + \cdots + \phi _{N-2} + \phi _{N-1})}\sin[\theta_1]\cos[\theta_2] \cdots \cos[\theta_{N-2}] \sin[\theta_{N-1}] \\
-\rme^{\rmi (\phi _{3} + \phi _{4} + \cdots + \phi _{N-2} + \phi _{N-1})}\sin[\theta_2]\cos[\theta_3] \cdots \cos[\theta_{N-2}] \sin[\theta_{N-1}] \\
\vdots\\
-\rme^{\rmi (\phi _{N-3} + \phi _{N-2} + \phi _{N-1})}\sin[\theta_{N-4}] \cos[\theta_{N-3}] \cos[\theta_{N-2}]\sin[\theta_{N-1}] \\
-\rme^{\rmi (\phi _{N-2} + \phi _{N-1})} \sin[\theta_{N-3}] \cos[\theta_{N-2}]\sin[\theta_{N-1}] \\
-\rme^{\rmi \phi_{N-1}} \sin[\theta_{N-2}]\sin[\theta_{N-1}] \\
\cos[\theta_{N-1}]
\end{array}\right],
\end{equation}
where $\varphi = \rme^{-\rmi \sqrt{\frac{2(N-1)}{N}} \, \phi_{(N^2+N-2)/2}}$ is an overall global phase.
More generally, since we are looking at the lowest weight state as the reference state, $SU(N)$ coherent states for $d$-dimensional systems can be easily shown to be equal to the $d$-th column of $U_{N}^{M}(\boldsymbol{\theta},\boldsymbol{\phi})$,
\begin{equation}
\label{TheCoherentStates-Generalized}
\ket{(\boldsymbol{\theta},\boldsymbol{\phi})_{N}^{M}} = U_{N}^{M}(\boldsymbol{\theta},\boldsymbol{\phi}) \ket{\psi_{0}} = [U_{N}^{M}(\boldsymbol{\theta},\boldsymbol{\phi})]_{d}.
\end{equation}
This is easy to see when $U_{N}^{M}(\boldsymbol{\theta},\boldsymbol{\phi})$ and $\ket{\psi_{0}}$ are represented in matrix form.  
The column vector of the lowest weight state has zero for all components apart from the bottom row element which is one.  
The only elements of the matrix $U_{N}^{M}(\boldsymbol{\theta},\boldsymbol{\phi})$ that are therefore relevant are those in the $d$-th column.  
If, on the other hand, we had chosen the highest weight state as the reference state, the resulting coherent state would be the first column of $U_{N}^{M}(\boldsymbol{\theta},\boldsymbol{\phi})$ i.\ e.\ $[U_{N}^{M}(\boldsymbol{\theta},\boldsymbol{\phi})]_{1}$.
Lastly, the coherent state~\eref{TheCoherentStates-Generalized} is equivalent to~\eref{coherent-state-atomic}, regardless of the value of $M$, for $N=2$~\cite{PhysRevA.6.2211,Perelomov,PerelomovB}, as well as more general coherent states for larger values of $N$ and $M$~\cite{Nemoto2000,Mathur2002}, if one makes the appropriate coordinate transformations on $\boldsymbol{\theta}$ and $\boldsymbol{\phi}$.

Lastly, by using the invariant volume element for the complex projective space in $N-1$ dimensions ($\rmd V_{\mathcal{C}P^{N-1}}$)~\cite{UandCPN}, 
\begin{eqnarray}
\label{eq:dvSUN}
& \rmd V_{\mathcal{C}P^{N-1}} = \biggl(\prod_{2 \leq y \leq N} K(y) \biggr) \rmd \theta_{N-1} \rmd \phi_{N-1} \ldots \rmd \theta_{1} \rmd \phi_{1}, \nonumber \\
& K(y) = 
\cases{
\sin[2\theta_{1}] & \qquad $y=2$, \\
\cos[\theta_{y-1}]^{2y-3}\sin[\theta_{y-1}] & \qquad $y\neq N$, \\
\cos[\theta_{N-1}]\sin[\theta_{N-1}]^{2N-3} & \qquad $y=N$, \\
}
\end{eqnarray} 
derived from~\eref{eq:suN} we have the following resolution of unity for $\ket{(\boldsymbol{\theta},\boldsymbol{\phi})_{N}^{M}}$,
\begin{equation}
\label{Normalization-1}
\fl
\frac{(N+M-1)!}{ 2\pi^{N-1} (M!)} \, \int_{\theta_1, \phi_1} \ldots \int_{\theta_{N-1},\phi_{N-1}}\ket{(\boldsymbol{\theta},\boldsymbol{\phi})_{N}^{M}}\bra{(\boldsymbol{\theta},\boldsymbol{\phi})_{N}^{M}} \, \rmd V_{\mathcal{C}P^{N-1}} = \Bid_{d}.
\end{equation}
We denote the identity matrix of size $d \times d$ by $\Bid_{d}$ and we are using the following integration ranges~\cite{UandCPN},
\begin{equation}
\label{eq:cpnranges1}
0 \le \theta_{i} \le \frac{\pi}{2} \text{ and } 0 \le \phi_{i} \le 2\pi \text{, for } 1 \le i \le N-1,
\end{equation}
to evaluate the integral.

As a final set of observations we note the following properties of our coherent states,
\begin{equation}
\label{Zeros}
\fl
\frac{(N+M-1)!}{ 2\pi^{N-1} (M!)} \,\int_{\theta_1, \phi_1} \ldots \int_{\theta_{N-1},\phi_{N-1}} \bra{(\boldsymbol{\theta},\boldsymbol{\phi})_{N}^{M}} \Lambda_{N,M}(k)\ket{(\boldsymbol{\theta},\boldsymbol{\phi})_{N}^{M}}\, \rmd V_{\mathcal{C}P^{N-1}} = 0
\end{equation}
for all $k$, and, as a special case when $M=1$,
\begin{eqnarray}
\label{Normalization-2}
\fl
\frac{N!}{2\pi^{N-1}} \, \int_{\theta_1, \phi_1} \ldots \int_{\theta_{N-1},\phi_{N-1}} &\bra{(\boldsymbol{\theta},\boldsymbol{\phi})_N^1} \Lambda_{N,1}(a) \ket{(\boldsymbol{\theta},\boldsymbol{\phi})_N^1} \nonumber \\
\fl
& \times \bra{(\boldsymbol{\theta},\boldsymbol{\phi})_N^1} \Lambda_{N,1}(b) \ket{(\boldsymbol{\theta},\boldsymbol{\phi})_N^1}  \; d V_{\mathcal{C}P^{N-1}} = \frac{2}{N+1} \; \delta_{ab},
\end{eqnarray}
where $\delta_{ab}$ is the Kronecker's delta. 

\section{$SU(N)$-Symmetric Distribution Functions}\label{Quasi-Straton}
Having defined our generalized coherent states, we can immediately generalize the Q-function to $SU(N)$ systems~\cite{0305-4470-34-10-309}.  
The Q-function of a density matrix $\rho$ may be written down as
\begin{equation}
\label{General-Q}
\mathrm{Q}(\boldsymbol{\theta},\boldsymbol{\phi}) = \bra{(\boldsymbol{\theta},\boldsymbol{\phi})_{N}^{M}} \rho \ket{(\boldsymbol{\theta},\boldsymbol{\phi})_{N}^{M}}.
\end{equation}
Here, $\Tr[\rho]=1$ and, using~\eref{eq:dvSUN},
\begin{equation}
\label{General-Q-Normalization}
\frac{(N+M-1)!}{ 2\pi^{N-1} (M!)} \, \int_{\theta_1, \phi_1} \ldots \int_{\theta_{N-1},\phi_{N-1}} \mathrm{Q}(\boldsymbol{\theta},\boldsymbol{\phi}) \, \rmd V_{\mathcal{C}P^{N-1}} = 1.
\end{equation}
Noting that 
\begin{equation}
\label{General-Q-CharacteristicFunction-1}
\bra{(\boldsymbol{\theta},\boldsymbol{\phi})_{N}^{M}} \rho \ket{(\boldsymbol{\theta},\boldsymbol{\phi})_{N}^{M}} = \Tr[\rho \cdot \ket{(\boldsymbol{\theta},\boldsymbol{\phi})_{N}^{M}}\bra{(\boldsymbol{\theta},\boldsymbol{\phi})_{N}^{M}}],
\end{equation}
we can see that we have the relation 
\begin{equation}
\label{General-Q-CharacteristicFunction-2}
\mathrm{Q}(\boldsymbol{\theta},\boldsymbol{\phi}) = \Tr[\rho \cdot \ket{(\boldsymbol{\theta},\boldsymbol{\phi})_{N}^{M}}\bra{(\boldsymbol{\theta},\boldsymbol{\phi})_{N}^{M}}].
\end{equation}

Equation~\eref{General-Q-CharacteristicFunction-2} is a special case of the more general Stratonovich-Weyl correspondence~\cite{Stratonovich56} that describes mappings between any Hilbert space operator $X$ and a characteristic function $f^{s}(\boldsymbol{\beta})$ on the classical phase space $\mathcal{X}$ ($\boldsymbol{\beta} \equiv \beta_{1}, \beta_{2} \ldots, \beta_{N}$ and $\boldsymbol{\beta} \in \mathcal{X}$) by
\begin{equation}
\label{Straton-Basic}
f^{s}(\boldsymbol{\beta}) = \Tr[X \cdot F^{s}(\boldsymbol{\beta})].
\end{equation}
This mapping is very useful in that it allows us to represent a density matrix as a distribution function in phase space.  
The quasi-distribution functions are information complete to their original density matrix.  
This means that one can reconstruct the density matrix from its quasi-distribution function, $f^{s}(\boldsymbol{\beta}) \mapsto X$, and the generating kernels $F^{s}(\boldsymbol{\beta})$ of $f^{s}(\boldsymbol{\beta})$, which we will build from a set of hermitian generators~\eref{TheLambdas}, satisfy
\begin{equation}
\label{Straton-Basic-Properties}
F^{s}(\boldsymbol{\beta}) = F^{s}(\boldsymbol{\beta})^\dagger \text{ and } 
\int_{\beta_1} \ldots \int_{\beta_N} F^{s}(\boldsymbol{\beta}) \, \rmd \nu(\boldsymbol{\beta})  = \Bid_N.
\end{equation}
Here $\Bid_N$ is the $N$-dimensional identity matrix and $s$ determines the type of distribution function being described~\cite{JPhysA.31.L9}:
\begin{eqnarray}
\label{Straton-Basic-Ordering}
f^{-1}(\boldsymbol{\beta}) & \rightarrow \text{P-function}, \nonumber \\
f^{+1}(\boldsymbol{\beta}) & \rightarrow \text{Q-function}, \nonumber \\
f^{0}(\boldsymbol{\beta}) & \rightarrow \text{Wigner function}. 
\end{eqnarray}
Because of the way the $f^{s}(\boldsymbol{\beta})$ are defined, they exhibit all the properties of the distribution functions they represent, however, from the application point of view, the value $s$ should be chosen dependent on the properties to be investigated.  
For instance, the Q-function is easy to calculate, but often obscures the quantum nature in the states.  
In these cases, the Wigner function tends to be preferred to represent the signature of quantum properties by interference fringes.
For finite-dimensional systems,~\eref{Straton-Basic} is a natural way to analyze the Wigner function, Q-function, and P-function, and we will build our generalized functions based on the various cases of~\eref{Straton-Basic}.
In detail,
\begin{equation}
\label{Straton-Basic-Mine}
f_{N,M,\rho}^{s}(\boldsymbol{\theta},\boldsymbol{\phi}) = \Tr[\rho \cdot F_{N,M}^{s}(\boldsymbol{\theta},\boldsymbol{\phi})],
\end{equation}
$F_{N,M}^{s}(\boldsymbol{\theta},\boldsymbol{\phi})$ is an operator of a $M$ representation of $SU(N)$,
and ($\boldsymbol{\theta},\boldsymbol{\phi}$) denotes the parameters from the coherent states given in~\eref{TheCoherentStates}. 

Following~\eref{General-Q-Normalization} and~\eref{Straton-Basic-Properties}, and using~\eref{eq:dvSUN} as our integral measure, we will demand that $f_{N,M,\rho}^{s}(\boldsymbol{\theta},\boldsymbol{\phi})$ and $F_{N,M}^{s}(\boldsymbol{\theta},\boldsymbol{\phi})$ satisfy
\begin{equation}
\label{Spin-M-Normalization-Main-General}
\frac{(N+M-1)!}{ 2\pi^{N-1} (M!)} \,\int_{\theta_1, \phi_1} \ldots \int_{\theta_{N-1},\phi_{N-1}} f_{N,M,\rho}^{s}(\boldsymbol{\theta},\boldsymbol{\phi})\, \rmd V_{\mathcal{C}P^{N-1}} = 1
\end{equation}
and
\begin{equation}
\label{Spin-M-Straton-Condition-2-Update-General}
\frac{(N+M-1)!}{ 2\pi^{N-1} (M!)} \, \int_{\theta_1, \phi_1} \ldots \int_{\theta_{N-1},\phi_{N-1}} F_{N,M}^{s}(\boldsymbol{\theta},\boldsymbol{\phi})\, \rmd V_{\mathcal{C}P^{N-1}} = \Bid_{d}.
\end{equation}
For example, if we define $F_{N,M}^{+1}(\boldsymbol{\theta},\boldsymbol{\phi}) = \ket{(\boldsymbol{\theta},\boldsymbol{\phi})_{N}^{M}}\bra{(\boldsymbol{\theta},\boldsymbol{\phi})_{N}^{M}}$ and $f_{N,M,\rho}^{+1}(\boldsymbol{\theta},\boldsymbol{\phi}) = \mathrm{Q}(\boldsymbol{\theta},\boldsymbol{\phi})$ then we can see that~\eref{Spin-M-Normalization-Main-General} and~\eref{Spin-M-Straton-Condition-2-Update-General} are satisfied by~\eref{Normalization-1} and~\eref{General-Q-Normalization}.
Lastly, to recover the density matrix, the relation
\begin{equation}
\label{Spin-M-Abs-Density-Recover-General}
\fl
\rho = \frac{(M+N-1)!}{ 2\pi^{N-1} (M!)} \, \int_{\theta_1, \phi_1} \ldots \int_{\theta_{N-1},\phi_{N-1}} f_{N,M,\rho}^{(+s)}(\boldsymbol{\theta},\boldsymbol{\phi}) F_{N,M}^{(-s)}(\boldsymbol{\theta},\boldsymbol{\phi})\, \rmd V_{\mathcal{C}P^{N-1}},
\end{equation}
has to be satisfied.

As we have seen in the beginning, the generalized Q-function is rather easy to construct, however the Wigner and P-functions are somewhat more complicated. 
Thus, we will first look into the $M=1$ case for arbitrary $SU(N)$ systems.

\subsection{Fundamental Representation}\label{Spin12}
As before, we start with the Q-function, which, in the fundamental representation, can be written as
\begin{equation}
\label{Spin12-Q-Straton}
f_{N,1,\rho}^{+1}(\boldsymbol{\theta},\boldsymbol{\phi}) = \Tr[\rho \cdot F_{N,1}^{+1}(\boldsymbol{\theta},\boldsymbol{\phi})],
\end{equation}
where
\begin{eqnarray}
\label{Spin12-Q-Formula}
\fl
F_{N,1}^{+1}(\boldsymbol{\theta},\boldsymbol{\phi}) &= \ket{(\boldsymbol{\theta},\boldsymbol{\phi})_N^1}\bra{(\boldsymbol{\theta},\boldsymbol{\phi})_N^1}, \nonumber \\
\fl
& =\frac{1}{N}\Bid_{N} + \frac{1}{2} \, \sum_{k=1}^{N^2-1} \bra{(\boldsymbol{\theta},\boldsymbol{\phi})_N^1}\Lambda_{N,1}(k) \ket{(\boldsymbol{\theta},\boldsymbol{\phi})_N^1} \Lambda_{N,1}(k).
\end{eqnarray} 
Now, in order for~\eref{Spin12-Q-Straton} to be useful, we need to be able to evaluate~\eref{Spin-M-Abs-Density-Recover-General}.
We can only do this if we know $F_{N,1}^{-1}(\boldsymbol{\theta},\boldsymbol{\phi})$, which is the generating kernel for the P-function.  
Demanding that we satisfy~\eref{Spin-M-Normalization-Main-General} and~\eref{Spin-M-Straton-Condition-2-Update-General}, we see that $F_{N,1}^{-1}(\boldsymbol{\theta},\boldsymbol{\phi})$ is 
\begin{equation}
\label{Spin12-P-Formula}
\fl
F_{N,1}^{-1}(\boldsymbol{\theta},\boldsymbol{\phi}) =\frac{1}{N}\Bid_{N} + \frac{N+1}{2} \, \sum_{k=1}^{N^2-1} \bra{(\boldsymbol{\theta},\boldsymbol{\phi})_N^1}\Lambda_{N,1}(k) \ket{(\boldsymbol{\theta},\boldsymbol{\phi})_N^1} \Lambda_{N,1}(k).
\end{equation} 
With~\eref{Spin12-P-Formula} done, the remaining function to generate is the Wigner function.  
Requiring that we again satisfy~\eref{Spin-M-Normalization-Main-General} and~\eref{Spin-M-Straton-Condition-2-Update-General}, we obtain for the Wigner function's generating kernel
\begin{equation}
\label{Spin12-Wigner-Formula}
\fl
F_{N,1}^{0}(\boldsymbol{\theta},\boldsymbol{\phi}) =\frac{1}{N}\Bid_{N} + \frac{\sqrt{N+1}}{2} \, \sum_{k=1}^{N^2-1} \bra{(\boldsymbol{\theta},\boldsymbol{\phi})_N^1}\Lambda_{N,1}(k) \ket{(\boldsymbol{\theta},\boldsymbol{\phi})_N^1} \Lambda_{N,1}(k).
\end{equation} 
Combining these results yields
\begin{eqnarray}
\label{Abs-SUN-Minea}
& F_{N,1}^{s}(\boldsymbol{\theta},\boldsymbol{\phi}) = \frac{1}{N}\Bid_{N} + \frac{\Omega(s)}{2} \, \sum_{k=1}^{N^2-1} \bra{(\boldsymbol{\theta},\boldsymbol{\phi})_N^1} \Lambda_{N,1}(k) \ket{(\boldsymbol{\theta},\boldsymbol{\phi})_N^1} \Lambda_{N,1}(k), \nonumber \\
& \Omega(s) =
\cases{
\sqrt{N+1} & \qquad $s=0$, \\
1 & \qquad $s=+1$, \\
N+1 & \qquad $s=-1$. \\
}
\end{eqnarray}
Substitution of~\eref{Abs-SUN-Minea} into~\eref{Straton-Basic-Mine}, with $M=1$, yields results that agree with existing Wigner, Q-, and P-functions for $N=2$ and $N=3$~\cite{Agarwal1981,Luis-495302}.
Lastly,~\eref{Abs-SUN-Minea} with~\eref{Straton-Basic-Mine} satisfies~\eref{Spin-M-Normalization-Main-General},~\eref{Spin-M-Straton-Condition-2-Update-General}, and~\eref{Spin-M-Abs-Density-Recover-General} for all three $s$ values.

Next, in the fundamental representation, a density matrix $\rho$ can be represented by~\cite{BillMunro1}
\begin{equation}
\label{Byrd-rho}
\rho = \frac{1}{N}\Bid_{N} + \sqrt{\frac{N-1}{2N}} \, \sum_{k=1}^{N^2-1}n_k \Lambda_{N,1}(k).
\end{equation}
For $N=2$, the set of all pure states are characterized by $\mathbf{n} \cdot \mathbf{n}$, $\mathbf{n} = \{n_k\}$, while for $N>2$, the set of all pure states are characterized by $\mathbf{n} \cdot \mathbf{n}$ and $\mathbf{n} \star \mathbf{n} = \mathbf{n}$ with the star product being defined in~\eref{Star}.
By substituting~\eref{Byrd-rho} into~\eref{Straton-Basic-Mine} and using~\eref{Abs-SUN-Minea}, as well as exploiting~\eref{LRules}, we get
\begin{equation}
\label{V-Luis-Substitution-simplified}
\fl
f_{N,1,\rho}^{s}(\boldsymbol{\theta},\boldsymbol{\phi})  = \frac{1}{N} + \sqrt{\frac{N-1}{2N}} \, \Omega(s) \, \sum_{k=1}^{N^2-1} \bra{(\boldsymbol{\theta},\boldsymbol{\phi})_N^1} \Lambda_{N,1}(k) \ket{(\boldsymbol{\theta},\boldsymbol{\phi})_N^1}n_k.
\end{equation}
To recover the density matrix from this function, we substitute~\eref{Abs-SUN-Minea} and~\eref{V-Luis-Substitution-simplified} into~\eref{Spin-M-Abs-Density-Recover-General}, as well as exploit~\eref{Zeros} and~\eref{Normalization-2}, to get (for the $s=0$ case)
\begin{eqnarray}
\label{General-C-Rho-Recovery}
\rho & = \frac{N!}{2\pi^{N-1}}\, \int_{\theta_1, \phi_1} \ldots \int_{\theta_{N-1},\phi_{N-1}} f_{N,1,\rho}^{0}(\boldsymbol{\theta},\boldsymbol{\phi}) F_{N,1}^{0}(\boldsymbol{\theta},\boldsymbol{\phi})\, \rmd V_{\mathcal{C}P^{N-1}}, \nonumber \\
& = \frac{1}{N} \Bid_{N} + \sqrt{\frac{N-1}{2N}} \,\sum_{j=1}^{N^2-1}n_k \Lambda_{N,1}(k).
\end{eqnarray} 
For the pure state case, the density matrix $\rho_{PS} = \ket{\psi}\bra{\psi}$ has the following $\mathbf{n}$ decomposition~\cite{MByrdKhaneja},
\begin{equation}
\label{ni}
n_k = \sqrt{\frac{N}{2(N-1)}} \, \bra{\psi}\Lambda_{N,1}(k)\ket{\psi}.
\end{equation}
Substitution of~\eref{ni} into~\eref{V-Luis-Substitution-simplified} and~\eref{General-C-Rho-Recovery} yields
\begin{equation}
\label{V-Luis-Substitution-simplified-PS}
\fl
f_{N,1,{\rho_{PS}}}^{s}(\boldsymbol{\theta},\boldsymbol{\phi}) = \frac{1}{N} + \frac{\Omega(s)}{2}\,\sum_{k=1}^{N^2-1} \bra{(\boldsymbol{\theta},\boldsymbol{\phi})_N^1} \Lambda_{N,1}(k) \ket{(\boldsymbol{\theta},\boldsymbol{\phi})_N^1}\bra{\psi} \Lambda_{N,1}(k) \ket{\psi},
\end{equation}
and
\begin{equation}
\label{General-C-Rho-Recovery-Final-PS}
\rho_{PS} = \frac{1}{N} \Bid_{N} + \frac{1}{2} \,\sum_{k=1}^{N^2-1} \bra{\psi} \Lambda_{N,1}(k) \ket{\psi}  \Lambda_{N,1}(k),
\end{equation}
which is equivalent to~\eref{Spin12-Q-Formula}.  
Similar calculations can be done for the $s=\pm1$ cases.

\subsection{Higher Dimensional Representations}\label{Spin22}
This problem cannot be simultaneously resolved for all three $s$ values by simply replacing $\ket{(\boldsymbol{\theta},\boldsymbol{\phi})_N^1}$ with more general spin coherent state representations from~\eref{TheCoherentStates}, or like those defined in Refs.~\cite{Nemoto2000,Mathur2002}, the full form of the kernels must be calculated.
To accomplish this, we start by generalizing~\eref{Abs-SUN-Minea} to generate a $M=2$ representation of $F_{N,M}^{s}(\boldsymbol{\theta},\boldsymbol{\phi})$,
\begin{equation}
\label{Spin-1-Abs-SUN-Mine}
F_{N,2}^{s}(\boldsymbol{\theta},\boldsymbol{\phi}) = \frac{1}{d}\Bid_{d} + \sum_{c=1}^2 \omega_{N,2}^{s}(c) \nu_{N,2}(\boldsymbol{\theta},\boldsymbol{\phi},c),
\end{equation}
where, using~\eref{TheOmega} and Section~\ref{GenCS},
\begin{eqnarray}
\label{Spin-1-Abs-SUN-W-Mine-2a}
\fl
\nu_{N,2}(\boldsymbol{\theta},\boldsymbol{\phi},c) & = \sum_{k=1}^{(b_{N,2-(c-1)})^2-1} \bra{(\boldsymbol{\theta},\boldsymbol{\phi})_N^2}\Lambda_{b_{N,2-(c-1)},c}(k)\ket{(\boldsymbol{\theta},\boldsymbol{\phi})_N^2}\Lambda_{b_{N,2-(c-1)},c}(k), 
\end{eqnarray}
and
\begin{eqnarray}
\label{Spin-1-Abs-SUN-W-Mine-2b}
\fl
\omega_{N,2}^{s}(c) & =
\cases{
\cases{
\frac{1}{2}\sqrt{b_{N+2,2}} & \, $c=1$, \\
\frac{1}{2 b_{N+2,1}}\biggl(\sqrt{\frac{b_{N+2,1}}{2}}-2 \,\omega_{N,2}^{0}(1) \biggr) &\, $c=2$, \\
} & \qquad $s=0$, \\
\cases{
\frac{1}{2} & \, $c=1$, \\
0 & \, $c=2$,  
} & \qquad $s=+1$, \\
\frac{(-1)^{c+1}}{2^c} \, b_{N+2,2-(c-1)} & \qquad $s=-1$. \\
}
\end{eqnarray}
It can be shown that~\eref{Spin-1-Abs-SUN-Mine} with~\eref{Straton-Basic-Mine} satisfies~\eref{Spin-M-Normalization-Main-General},~\eref{Spin-M-Straton-Condition-2-Update-General}, and~\eref{Spin-M-Abs-Density-Recover-General} for all three $s$ values.

As an example, we note that for a $M=2$ $SU(2)$ system, when $s=0$, our new $F_{N,2}^{s}(\boldsymbol{\theta},\boldsymbol{\phi})$ gives us
\begin{eqnarray}
\label{Spin-1-Abs-21}
\fl
&F_{2,2}^{0}(\boldsymbol{\theta},\boldsymbol{\phi}) = \frac{1}{d}\Bid_{d} + \sum_{c=1}^2 \omega_{2,2}^{0}(c) \nu_{2,2}(\boldsymbol{\theta},\boldsymbol{\phi},c), \nonumber \\
\fl
& = \frac{1}{3}\Bid_{3} +\omega_{2,2}^{0}(1) \sum_{k=1}^{8} \bra{(\boldsymbol{\theta},\boldsymbol{\phi})_2^2}\Lambda_{3,1}(k)\ket{(\boldsymbol{\theta},\boldsymbol{\phi})_2^2}\Lambda_{3,1}(k)  \\
\fl
& \quad + \omega_{2,2}^{0}(2) \sum_{k=1}^{3} \bra{(\boldsymbol{\theta},\boldsymbol{\phi})_2^2}\Lambda_{2,2}(k)\ket{(\boldsymbol{\theta},\boldsymbol{\phi})_2^2}\Lambda_{2,2}(k), \nonumber \\
\fl
& = \frac{1}{3}\Bid_{3} + \frac{\sqrt{10}}{2} \sum_{k=1}^{8} \bra{(\boldsymbol{\theta},\boldsymbol{\phi})_2^2}\lambda_{k} \ket{(\boldsymbol{\theta},\boldsymbol{\phi})_2^2}\lambda_{k} + \frac{(\sqrt{2} - \sqrt{10})}{8} \sum_{k=1}^{3}\bra{(\boldsymbol{\theta},\boldsymbol{\phi})_2^2} J_{k} \ket{(\boldsymbol{\theta},\boldsymbol{\phi})_2^2} J_{k}. \nonumber
\end{eqnarray}
Here we have again used~\eref{TheOmega} and Section~\ref{GenCS} to recognize that $\Lambda_{2,2}(K) \equiv J_{k}$, the spin-$1$ representation of the $SU(2)$ Pauli spin matrices, and that $\Lambda_{3,1}(k) \equiv \lambda_k$, the standard Gell-Mann matrices for $SU(3)$.
Lastly, evaluating~\eref{TheCoherentStates} for $N,M=2$ yields
\begin{equation}
\label{TheSU2-Spin1-CS}
\ket{(\boldsymbol{\theta},\boldsymbol{\phi})_2^2} = \rme^{-2 i \phi_2} 
\left[\begin{array}{c}
\rme^{2 \rmi \phi _1} \sin[\theta_1]^2 \\
\frac{\sin[2 \theta_1]}{\sqrt{2}} \\
\rme^{-2 \rmi \phi_1} \cos[\theta_1]^2
\end{array}\right].
\end{equation}
It is easy to verify that~\eref{Spin-1-Abs-21} yields an equivalent operator for the Wigner function as that given in Refs.~\cite{Agarwal1981,KlimovSpanish,Klimov055303} and, using~\eref{Straton-Basic-Mine}, gives equivalent Wigner functions as that given in~\cite{PhysRevA.49.4101}.
Furthermore, a similar calculation for the $s=\pm1$ cases yield equivalent operators as that given in Ref.~\cite{Agarwal1981} for the Q-function and P-function.
Therefore, despite the apparent differences between expressions,~\eref{Spin-1-Abs-SUN-Mine} is the appropriate generalization of~\eref{Abs-SUN-Minea} to $M=2$.

\section{Discussion and Conclusion}\label{Properties}
In this section, we mainly discuss two issues: the graphical representation of states of $N$-level systems using our distribution functions and the correspondence relation between the various distribution functions.  
First, we show some examples of the graphical representation for a $SU(4)$ system by considering a Werner state~\cite{Werner} of two qubits,
\begin{equation}
\label{SU4-Density}
\rho_{\text{Werner}} = \frac{1}{4}
\left( \begin{array}{cccc}
1-\gamma & 0 & 0 & 0 \\
0 & 1+\gamma & -2\gamma & 0 \\
0 & -2\gamma & 1+\gamma & 0 \\
0 & 0 & 0 & 1-\gamma
\end{array} \right). 
\end{equation} 
Here the parameter $\gamma$ defines the purity of the state.
In particular $\gamma=1$ corresponds a pure state whereas $\gamma=0$ gives the completely mixed state.
Evaluating $\rho_{\text{Werner}}$ via~\eref{Byrd-rho} gives us
\begin{equation}
\label{Byrd-rho-Werner-n}
n_3=-\frac{\gamma}{\sqrt{6}}, \, n_6=-\gamma\sqrt{\frac{2}{3}}, \, n_8=-\frac{\gamma}{3 \sqrt{2}}, \text{ and } n_{15}=\frac{\gamma}{3},
\end{equation}
which can be substituted into~\eref{V-Luis-Substitution-simplified} to yield
\begin{eqnarray}
\label{SU4-Werner-CF} 
\fl
f_{4,1,\rho_{\text{Werner}}}^{s}(\boldsymbol{\theta},\boldsymbol{\phi}) &= \frac{1}{24} \biggl\{6 - \gamma \biggl(2 + 4 \cos[2 \theta_3] - (1-6 \cos[2 \theta_1] \cos[\theta_2]^2 \nonumber \\
\fl
&\,\,- 3\cos[2 \theta_2] - 12 \cos[\phi_1-\phi_2] \sin[\theta_1]\sin[2 \theta_2]) \sin[\theta_3]^2\biggr) \Omega(s)\biggr\}.
\end{eqnarray}
Here, $\Omega(s)$ is as defined in~\eref{Abs-SUN-Minea} with $N=4$, i.\ e.\ $\Omega(0) = \sqrt{5}$, $\Omega(+1) = 1$, and $\Omega(-1) = 5$.

\begin{figure}[ht]
\begin{center}
\includegraphics[scale=0.55]{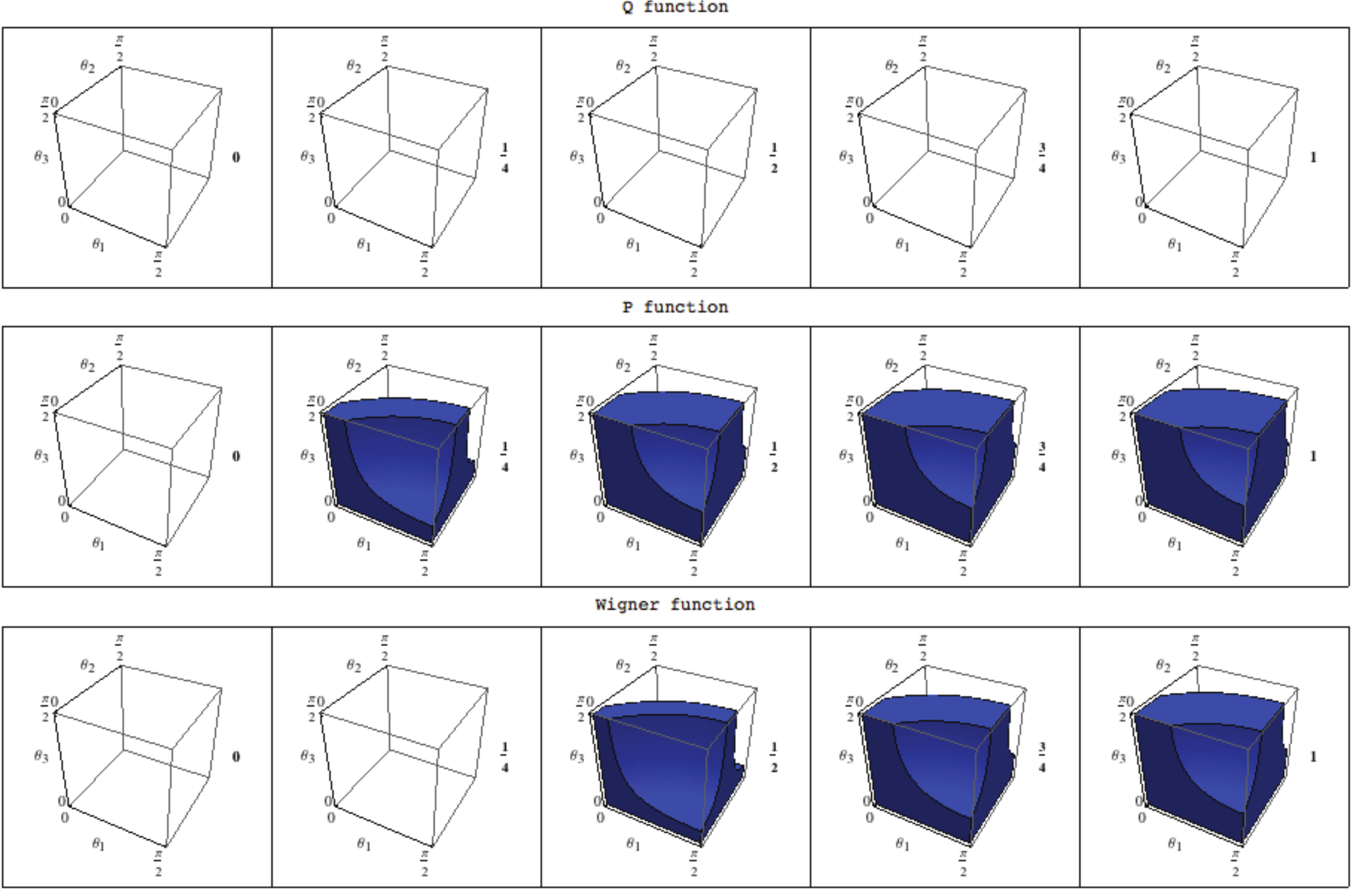}
\end{center}
\caption{Graphical representation of the Q (top), P (middle), and Wigner (bottom) functions from~\eref{positivenegative-SU4-Werner-CF} for $\phi_1, \phi_2=0$ and various values of $\gamma$.  
The three plots at far left show the case of the completely mixed state with $\gamma=0$ and the far right ones plot the pure state with $\gamma=1$. The value of $\gamma$ for each plot is at right of the graphic.
The colored areas represent those values of $\theta_1$, $\theta_2$, and $\theta_3$ where the corresponding function is negative.}
\label{fig2a}
\end{figure}

\begin{figure}[ht]
\begin{center}
\includegraphics[scale=0.55]{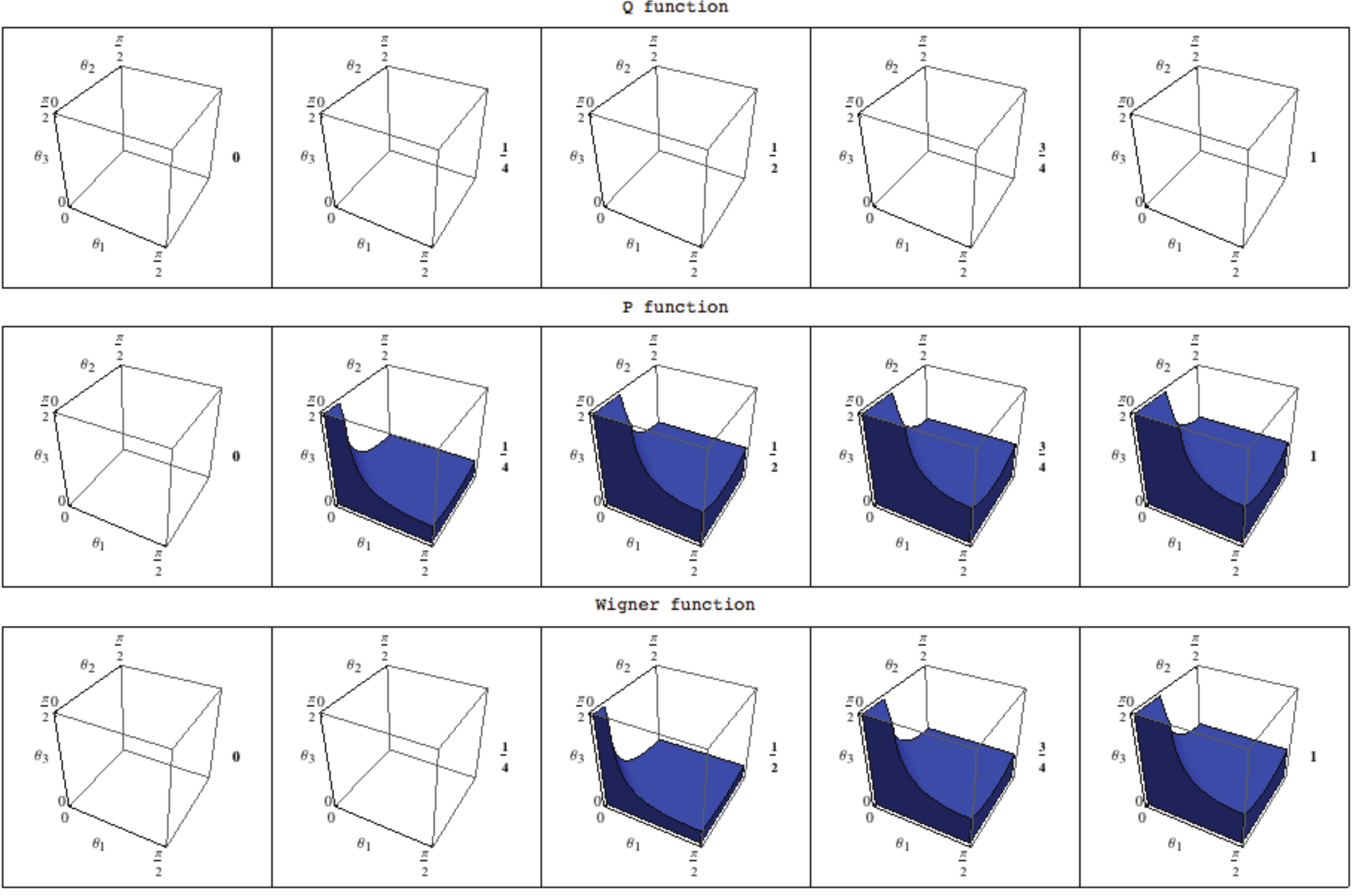}
\end{center}
\caption{Graphical representation of the Q (top), P (middle), and Wigner (bottom) functions from~\eref{positivenegative-SU4-Werner-CF} for $\phi_1=\pi,\, \phi_2=0$ and various values of $\gamma$.
The three plots at far left show the case of the completely mixed state with $\gamma=0$ and the far right ones plot the pure state with $\gamma=1$. The value of $\gamma$ for each plot is at right of the graphic.
The colored areas represent those values of $\theta_1$, $\theta_2$, and $\theta_3$ where the corresponding function is negative.}
\label{fig2b}
\end{figure}

The function given in~\eref{SU4-Werner-CF} is expressed in a five-parameter space, thus it is not easy to represent its entire property in one figure.  
So, we take cross sections of the function.  
To do this efficiently, we look at the element of the phases $\phi_{1}$ and $\phi_{2}$ in~\eref{SU4-Werner-CF}, that is $12 \cos[\phi_{1}-\phi_{2}]$.  
This element is effectively one parameter ($\phi_{1}-\phi_{2}$), so we can set it to have two extreme cases: $\phi_{1}, \phi_{2}=0$ and $\phi_{1}=\pi, \, \phi_{2}=0$.
Doing this gives
\begin{eqnarray}
\label{positivenegative-SU4-Werner-CF}
\fl
f_{4,1,\rho_{\text{Werner}}}^{s}(\boldsymbol{\theta},\boldsymbol{\phi}) &= \frac{1}{24} \biggl\{6 - \gamma \biggl(2 + 4 \cos[2 \theta_3] - (1-6 \cos[2 \theta_1] \cos[\theta_2]^2 \nonumber \\
\fl
&\,\,- 3\cos[2 \theta_2] \mp 12 \sin[\theta_1]\sin[2 \theta_2]) \sin[\theta_3]^2\biggr) \Omega(s)\biggr\}.
\end{eqnarray}
Figures~\ref{fig2a} and~\ref{fig2b} show the parameter regions where these quasi-distribution functions exhibit negative values for both cases $\phi_{1}, \phi_{2}=0$ and $\phi_{1}=\pi, \, \phi_{2}=0$ respectively.  
As we expect, the Q-function does not yield negative values in the entire parameter regime, however when the states are pure enough, the P-function and the Wigner function can be negative.  
When the purity of the Werner state, indicated by $\gamma$, becomes small enough, i.\ e.\ when the state is more mixed, the parameter regime for the negative value disappears.  
In fact, the P-function can be negative when $p \geq 1/4$, while the Wigner function shows a negative region when $p \geq 1/2$.  
Such negativity in these quasi-distribution functions is often considered evidence of quantum nature in the states, however, our results indicates such a simple explanation does not apply to $SU(N)$ systems.

There are of course other ways to reduce the dimensionality of the parameter space.  
The optimality of a representation is dependent on the properties we are after.  
Taking a cross section is the easiest way to generate a graphical representation to discern rough properties of the state.  

Next, we briefly discuss the correspondence between different distribution functions.  
As we mentioned before, the concrete expression of the various distribution functions depends on the parametrization of the $SU(N)$ group operators.
In this paper, we employed the parametrization given in~\cite{Tilma2}, which is an extension of the ones used for $SU(2)$ and $SU(3)$.
This parametrization gives us a way to write out the Wigner function, Q-function, and P-function that, in the $M=1$ and $M=2$ cases, makes their correspondence easy to see.
Furthermore, it shows us how more general $M$ representations should be related.

To begin, for the $M=1$ case, we start by making the following redefinition of~\eref{Abs-SUN-Minea},
\begin{equation}
\label{Abs-SUN-Minea-Redefine}
\tilde{F}_{N,1}^{s}(\boldsymbol{\theta},\boldsymbol{\phi}) = F_{N,1}^{s}(\boldsymbol{\theta},\boldsymbol{\phi}) - \frac{1}{N}\Bid_{N},
\end{equation}
such that
\begin{equation}
\label{Abs-SUN-Minea-Redefine-Part2}
\tilde{F}_{N,1}^{s'}(\boldsymbol{\theta},\boldsymbol{\phi}) = \frac{\Omega(s')}{\Omega(s)} \, \tilde{F}_{N,1}^{s}(\boldsymbol{\theta},\boldsymbol{\phi}) 
\end{equation}
is true.
This can only be done because our $\Omega(s)$ terms in~\eref{Abs-SUN-Minea} are all positive definite.
Using~\eref{Abs-SUN-Minea-Redefine} and~\eref{Abs-SUN-Minea-Redefine-Part2} we therefore get
\begin{eqnarray}
\label{Abs-SUN-Minea-Conversions}
F_{N,1}^{s'}(\boldsymbol{\theta},\boldsymbol{\phi}) & = \frac{\Omega(s')}{\Omega(s)} \, \tilde{F}_{N,1}^{s}(\boldsymbol{\theta},\boldsymbol{\phi}) + \frac{1}{N}\Bid_{N}, \nonumber \\
& = \frac{\Omega(s')}{\Omega(s)} \, \biggl(F_{N,1}^{s}(\boldsymbol{\theta},\boldsymbol{\phi}) - \frac{1}{N}\Bid_{N} \biggr) + \frac{1}{N}\Bid_{N}, \nonumber \\
& = \frac{\Omega(s')}{\Omega(s)} \, F_{N,1}^{s}(\boldsymbol{\theta},\boldsymbol{\phi}) + \biggl( 1 - \frac{\Omega(s')}{\Omega(s)} \biggr)\frac{1}{N}\Bid_{N}.
\end{eqnarray}
From this we can see how we can convert between the various $F_{N,1}^{s}(\boldsymbol{\theta},\boldsymbol{\phi})$ operators and, through~\eref{Straton-Basic-Mine}, the various $f_{N,1,\rho}^{s}(\boldsymbol{\theta},\boldsymbol{\phi})$.

A similar procedure can be done for the $M=2$ case.
In detail, following~\eref{Abs-SUN-Minea-Redefine} we redefine~\eref{Spin-1-Abs-SUN-Mine} to give us
\begin{equation}
\label{Spin-1-Abs-SUN-Mine-Redefine}
\tilde{F}_{N,2}^{s}(\boldsymbol{\theta},\boldsymbol{\phi}) = F_{N,2}^{s}(\boldsymbol{\theta},\boldsymbol{\phi}) - \frac{1}{d}\Bid_{d}.
\end{equation}
By construction, $\tilde{F}_{N,2}^{s}(\boldsymbol{\theta},\boldsymbol{\phi})$ is non-singular and of rank $d$.
It is therefore invertible, allowing us to have
\begin{equation}
\label{Spin-1-Abs-SUN-Mine-Redefine-Part1}
\tilde{F}_{N,2}^{s'}(\boldsymbol{\theta},\boldsymbol{\phi}) = \tilde{F}_{N,2}^{s'}(\boldsymbol{\theta},\boldsymbol{\phi}) \cdot \tilde{F}_{N,2}^{s}(\boldsymbol{\theta},\boldsymbol{\phi})^{-1} \cdot \tilde{F}_{N,2}^{s}(\boldsymbol{\theta},\boldsymbol{\phi}).
\end{equation}
This allows us to state (via~\eref{Spin-1-Abs-SUN-Mine-Redefine}) that 
\begin{eqnarray}
\label{Spin-1-Abs-SUN-Mine-Conversions}
\fl
\tilde{F}_{N,2}^{s'}(\boldsymbol{\theta},\boldsymbol{\phi}) & + \frac{1}{d}\Bid_{d} & =  \tilde{F}_{N,2}^{s'}(\boldsymbol{\theta},\boldsymbol{\phi}) \cdot \tilde{F}_{N,2}^{s}(\boldsymbol{\theta},\boldsymbol{\phi})^{-1} \cdot \tilde{F}_{N,2}^{s}(\boldsymbol{\theta},\boldsymbol{\phi}) + \frac{1}{d}\Bid_{d}, \nonumber \\
\fl
& F_{N,2}^{s'}(\boldsymbol{\theta},\boldsymbol{\phi}) & = \tilde{F}_{N,2}^{s'}(\boldsymbol{\theta},\boldsymbol{\phi}) \cdot \tilde{F}_{N,2}^{s}(\boldsymbol{\theta},\boldsymbol{\phi})^{-1} \biggl(F_{N,2}^{s}(\boldsymbol{\theta},\boldsymbol{\phi}) - \frac{1}{d}\Bid_{d} \biggr) + \frac{1}{d}\Bid_{d}, \nonumber \\
\fl
&& = \Upsilon(s',s,2)\cdot F_{N,2}^{s}(\boldsymbol{\theta},\boldsymbol{\phi}) + (1 - \Upsilon(s',s,2)) \cdot \frac{1}{d}\Bid_{d},
\end{eqnarray}
where we have made the following definition: $\Upsilon(s',s,2) = \tilde{F}_{N,2}^{s'}(\boldsymbol{\theta},\boldsymbol{\phi}) \cdot \tilde{F}_{N,2}^{s}(\boldsymbol{\theta},\boldsymbol{\phi})^{-1}$. 
Conversion between the various $f_{N,2,\rho}^{s}(\boldsymbol{\theta},\boldsymbol{\phi})$ via~\eref{Straton-Basic-Mine} is now straightforward.

In general, we can see that the transformation sequence between the various functions in the general $M$ case is equivalent to~\eref{Spin-1-Abs-SUN-Mine-Conversions} if we make the following definition: $\Upsilon(s',s,M) = \tilde{F}_{N,M}^{s'}(\boldsymbol{\theta},\boldsymbol{\phi}) \cdot \tilde{F}_{N,M}^{s}(\boldsymbol{\theta},\boldsymbol{\phi})^{-1}$. 
For example, when $M=1$, $\Upsilon(s',s,1)$ reduces to $\Omega(s') / \Omega(s)$ as expected.

To conclude, in this paper we have given an explicit set of $SU(N)$-symmetric functions that represent finite-dimensional versions of the Wigner function, Q-function and P-function by using generalized coherent state.  
In the case of the general $M$ $SU(2)$ and $M=1$ $SU(3)$ representations, these functions are equivalent to previously derived finite-dimensional Wigner, Q-, and P-functions with an appropriate parameter change~\cite{Agarwal1981,Wolf6247,Luis-052112,Luis-495302,KlimovSpanish,Klimov055303}.  
The quasi-probability distribution functions in this paper have been generalized to a higher quantum number $M=2$.  
Such quasi-probability distribution functions may also have some benefits to characterize qubit-qunat systems.  
These hybrid systems are becoming extensively investigated in the context of quantum information devices. 
For more complex systems, there are possibilities to further generalize the formula to an arbitrary $M$. 
However, the analysis has showed that the process is not as straightforward as the $SU(2)$ case~\cite{PhysRevA.49.4101} and further work will be necessary to complete the generalization.  

\section*{Acknowledgements}\label{Ack}
We would like to thank Jon Dowling, Bill Munro, and Peter Turner for helpful discussions. 
This work was partly supported by NICT and JSPS.

\appendix
\section{}\label{Lambda}
\setcounter{section}{1}
The $\{ \Lambda_{N,M}(k) \}$ matrices given in~\eref{TheLambdas} are a subset of the corresponding Lie algebra of $SU(N)$; a set of Hermitian, traceless matrices of size $d \times d$ that are defined in the following way~\cite{Nemoto2000}:
\begin{enumerate}
\item{Define a general basis $\ket{m_1,m_2,\ldots,m_N}$ where $M=\sum_{k=1}^{N}m_k$, $m_{k} \in \mathbb{Z}$, and $M \in \mathbb{Z^{+}}$.}
\item{Define the following three operators:
\begin{equation*}
\label{General-Lambda-Jab}
\fl
J_b^a \ket{m_1,\ldots,m_a,m_b,\ldots,m_N} = \sqrt{(m_a+1)m_b} \, \ket{m_1,\ldots,m_a+1,m_b-1,\ldots,m_N} \nonumber
\end{equation*}
for $1 \leq a < b \leq N$,
\begin{equation*}
\label{General-Lambda-Jba}
\fl
J_b^a \ket{m_1,\ldots,m_b,m_a,\ldots,m_N} = \sqrt{m_a(m_b+1)} \, \ket{m_1,\ldots,m_a-1,m_b+1,\ldots,m_N} \nonumber 
\end{equation*}
for $1 \leq b < a \leq N$, and
\begin{equation*}
\label{General-Lambda-C}
\fl
J_c^c \ket{m_1,\ldots,m_c,\ldots,m_N} = \sqrt{\frac{2}{c(c+1)}} \, \biggl(\sum_{k=1}^c m_k-cm_{c+1}\biggr)\ket{m_1,\ldots,m_c,\ldots,m_N} \nonumber
\end{equation*}
for $1 \leq c \leq N-1$.
}
\item{Using the basis given in (i) and the operators given in (ii), define the following matrices:
\begin{eqnarray}
\label{General-Lambda-Final}
&\Lambda_{N,M}^{\{1 \}}(a,b) \equiv J_b^a + J_a^b, \nonumber \\ 
&\Lambda_{N,M}^{\{2 \}}(a,b) \equiv -\rmi (J_b^a - J_a^b ), \nonumber \\ 
&\Lambda_{N,M}^{\{3 \}}((c+1)^2-1) \equiv J_c^c.
\end{eqnarray}
for $a,b = 1,2,3,\ldots,N;\, a < b$ and $c=1,2,\ldots,N-1$.
}
\item{Combine the three matrices given in~\eref{General-Lambda-Final} to yield the set $\{ \Lambda_{N,M}(k) \}$ where $k=1,2,\ldots,N^2-1$.
}
\end{enumerate}
In general, our lambda matrices can be used to define the $M=1$ star product,
\begin{equation}
\label{Star}
(\mathbf{x} \star \mathbf{y})_k = \sqrt{\frac{N(N-1)}{2(N-2)^2}} \, \Tr[\{\Lambda_{N,1}(i), \Lambda_{N,1}(j)\} \cdot \Lambda_{N,1}(k)] x_i y_j,
\end{equation}
as well as be shown to satisfy
\begin{eqnarray}
\label{LRules}
& \Tr[\Lambda_{N,M}(i) \cdot \Lambda_{N,M}(j)] = \frac{2M}{N+1} b_{N+1,M}\delta_{ij}, \nonumber \\
& [ \Lambda_{N,M}(i) , \Lambda_{N,M}(j) ] = c \times f_{ijk} \, \Lambda_{N,M}(k), \nonumber \\
& f_{ijk} = \frac{1}{2c} \times \Tr[ [ \Lambda_{N,M}(i) , \Lambda_{N,M}(j) ] \cdot \Lambda_{N,M}(k) ],
\end{eqnarray}
thus forming a basis for the corresponding vector space, and a representation of the spin generators of $SU(N)$.
For example, when $M=1$ and $c=2 \rmi$,~\eref{General-Lambda-Final} and~\eref{LRules} reproduce the form, and properties, of the generalized Gell-Mann matrices~\cite{Greiner,Georgi}.

%
%

\end{document}